\documentclass[12pt,preprint]{aastex}

\shortauthors{L. \ Bouchet et al.}
\shorttitle
{SPI/INTEGRAL observations of the Galactic central radian}

\begin{document}

\title{SPI/INTEGRAL observation of the Galactic central radian:  
contribution of discrete sources and implication for the diffuse emission \footnote{
Based on observations with INTEGRAL, an ESA project with instruments and science data centre
funded by ESA member states (especially the PI countries: Denmark, France, Germany, Italy, 
Spain, and Switzerland), Czech Republic and Poland with participation of Russia and USA.} }

\author{L. Bouchet, J.P.Roques, P. Mandrou}
\affil{CESR--CNRS, 9 Av. du Colonel Roche, 31028 Toulouse Cedex~04, France}
\and
\author{A. Strong, R. Diehl} 
\affil{Max-Planck-Institut f\"ur extraterrestrische Physik, Postfach 1603, 85740 Garching, Germany}

\author{F. Lebrun}
\affil{DSM/DAPNIA/SAp, CEA-Saclay, 91191 Gif-sur-Yvette, France} 

\author{R. Terrier}
\affil{APC, UMR 7164, 11 place M. Berthelot, 75005 Paris, France} 

\author{\it Received 2005 May 14 ; accepted 2005, August 19}


\begin{abstract}

The INTEGRAL observatory has been performing a deep survey of the Galactic central radian 
since 2003, with the goal of both extracting a catalog of sources and
gaining insight into the Galactic diffuse emission.
This paper concentrates on the estimation of the total point sources emission contribution. It is now 
clear that unresolved point sources contribute to the observed diffuse emission; 
the increasing sensitivity of instruments with time has lead to a steady decrease in estimates of this
``diffuse emission''. 

We have analysed the first year data obtained with the spectrometer and imager
SPI on board INTEGRAL.    
First, a catalog of 63 hard X-ray sources detected,
time-averaged, during our 2003 Galactic plane survey, is derived. Second, after extracting the spectra of the 
sources detected by SPI, their combined contribution is compared to the total 
(resolved and unresolved) emission from the Galactic ridge. The data analysis is complex: it requires us
to split the total emission into several components, as discrete sources and diffuse emission are
superimposed in SPI data.
The main result is that point source emission dominates in the hard X-ray/soft 
$\gamma$-ray domain, and contributes around 90 \% of the total emission around 100 keV, while
above 250 keV, diffuse electron-positron annihilation, through its three-photon positronium continuum
 with a positronium fraction
$\sim$ 0.97 and the 511 keV electron-positron line, dominates over the sources.

\end{abstract}

\keywords{Galaxy: general-- gamma rays: observations -- surveys -- cosmic rays}


\section{Introduction}

 
Observations carried out for more than three decades indicate that the 
total hard X-ray/soft $\gamma$-ray emission from the Galaxy results from the superposition 
of multiple physical processes 
whose relative contribution depends on the energy.

The spectrum of this emission is reasonably well measured and understood above
$\sim$  1 MeV from OSSE and COMPTEL (Kinzer et al., 1999, Strong et al., 1994, 2004a). 
At energies above $\sim$ 100 MeV, the dominant emission process is the decay of $\pi^0$ mesons 
produced in the interaction of cosmic-ray nucleons with the interstellar matter (Bertsch et al., 1993). 
Between ~1 and 70 MeV, electron bremsstrahlung and inverse Compton scattering are expected 
to dominate over discrete source emission (Sacher \& Sch\"onfelder, 1984, Skibo et al., 1993, Strong et al. 2004a ).

In the low-energy domain, the spectrum has been measured with RXTE in the 10-35 keV energy band 
(Valinia \& Marshall, 1998) and with Ginga in the 3-16 keV band (Yamasaki et al., 1997). More
recently ASCA (Sugizaki et al., 2001) observations in the 0.7-10 keV band,
 and Chandra observations (Ebisawa et al., 2001) in
the 2-10 keV band, show that there is a genuine diffuse soft X-ray component,
and  this result is confirmed by XMM-Newton measurements 
(Hands et al., 2004) in the 2-10 keV band. 

However, in the hard X-ray/soft $\gamma$-ray band (50-500 keV), the situation is
more complicated as multiple components are believed to contribute to the 
total emission. These include discrete sources, the positron annihilation line, 
three-photon positronium continuum radiation and a soft $\gamma$-ray component 
corresponding to the diffuse continuum emission induced by cosmic-ray interactions (CR diffuse).

Unfortunately, measurements of the Galactic diffuse emission in the hard 
X-ray/soft $\gamma$-ray band and its interpretation are inherently
difficult because of the presence of numerous hard X-ray
discrete sources in the Galactic plane. Moreover, generally, hard 
X-ray/soft $\gamma$-ray instruments either have large fields of view and no imaging 
capabilities or have imaging capability but no sensitivity to extended emission. As a result,
discrimination between diffuse emission and point sources 
remains a difficult task. For this reason, simultaneous multiple-instrument observations
had been performed, with coordinated observations of the Galactic 
Center region with OSSE/CGRO (Kurfess et al., 1991) and the imaging instrument 
SIGMA/GRANAT (Paul et al, 1991).
The main results were that, in the hard X-ray/soft $\gamma$-ray regime,
the spatial distribution of ``diffuse emission'' is
broad and relatively flat in longitude, with an extension in latitude of 
$\sim$ $5.5^\circ$ width, while a few discrete sources contribute at 
least 50$\%$ of the total emission (Purcell et al., 1996). However SIGMA had a sensitivity of about 25 mcrab 
(2$\sigma$) for a typical 24 hours observation. As a result, weak sources 
escaped detection in its survey, and the unresolved emission was still  
contaminated significantly by discrete sources.
The extension of the spectral shape of the Galactic diffuse emission  from 
the hard X-ray to $\gamma$-ray regime, and how much of the emission was due to 
discrete sources remained to be precisely determined.  
Further attempts to derive the diffuse
emission characteristics followed (Kinzer et al. 1999, 2001; Boggs
et al., 1999; Valinia et al., 2000).

Theoretical studies were unable to explain even 50\% of the Galactic emission as originating
in the interstellar medium (ISM).
Two main processes can lead to an interstellar soft $\gamma$-ray emission. 
The first one is inverse Compton scattering of high energy (GeV) cosmic ray 
electrons on the ambient photons field; but these electrons would also produce radio synchrotron emission 
in the Galactic magnetic field at a level much higher than  one actually 
observed. The second process is bremsstrahlung of a population 
of electrons of a few hundred keV, radiating through interactions
with interstellar gas. Because these electrons will lose their  energy
through ionization and Coulomb collisions, the total power required to 
compensate for these energy losses is of the order of $10^{41} - 10^{43}$ erg$~$s$^{-1}$.
This power, comparable or higher than that of cosmic ray protons, would affect 
interstellar-medium ionization equilibrium and give rise to excessive 
dissociation of interstellar molecules. 
A possible interstellar process has been proposed by Dogiel et al. (2002).
Alternatively, a dozen  point sources with intensities around 25 mCrab (the SIGMA  2$\sigma$ 
sensitivity) could account for most of the remaining diffuse emission.
This demonstrates the need for highly sensitive imaging instruments
for a precise determination of the diffuse Galactic emission.\\
Actually, a new vision of the hard X ray sky is provided by INTEGRAL with the detection
of a significant amount of new sources and the discovery  of a new class of objects:
the highly absorbed sources (Dean et al., 2005) may represent 20\% of the total number of sources,
but their contribution could not be evaluated by X-ray survey, with an obvious implication for the
diffuse emission.

A recent result based on ISGRI/IBIS/INTEGRAL data (Hereafter ISGRI) shows that known binary 
sources account for the main part of the total Galactic emission (from 86\% to 74\%) 
from 20 keV to 220 keV (Lebrun et al., 2004, Terrier et al. 2004).
However the limited sensitivity of ISGRI, above 200 keV, does not allow an extension of this 
study at higher energies. 
Early study of the diffuse continuum emission using SPI/INTEGRAL (hereafter SPI) shows that
diffuse emission dominates above $\sim$ 200 keV (Strong et al. 2003) and that 
the ratio of diffuse emission to total emission vary from $10\%$ to $100\%$ 
along SPI energy domain (Strong et al. 2004b).

SPI (see next section) has a moderate angular resolution
over a large field, thus is sensitive to
extended sources  as well as to point sources imaging. SPI can thus be used in a self consistent 
way to measure both the Galactic
diffuse and discrete source emission, avoiding the complication due 
to different regions observed, different instrument responses, different
models assumed and diverse ways in which results are presented (per radian,
per Field-Of-View etc.) when comparisons with other instruments are made.

We present the catalog of the sources detected with SPI as well as
their global spectral emission. An early SPI sources catalog can be found in
Bouchet et al. (2004). Moreover a broadband spectral analysis
including point sources and diffuse emission components (annihilation,
positronium and CR diffuse continuum) has been performed in a self consistent way, i.e using
the same instrument for both sources detection and diffuse emission estimate.
We draw attention to an independent, complementary study on this topic by Strong et al. (2005).

\section{Instrument}


The spectrometer SPI (Vedrenne et al., 2003) is one of the two main instruments onboard 
ESA's INTEGRAL (INTErnational Gamma-Ray Astrophysics Laboratory) observatory 
launched from Baikonour, Kazakhstan, on 2002 October 17. 

It consists of an array of 19 actively cooled high resolution
Germanium (Ge) detectors with an area of 508 cm$^{2}$ and a thickness of 7 cm. 
It is surrounded by a 5-cm thick BGO shield. The detectors cover the 20 keV - 8 MeV energy range 
with an energy resolution ranging from  2 to 8 keV as 
a function of energy. In addition to its spectroscopic capability, SPI can 
image the sky with a spatial resolution of $2.6^\circ$ over a field 
of view of $30^\circ$, thanks to its coded mask. Despite such a modest 
angular resolution, it is possible to locate intense sources with an accuracy of few arc minutes 
(Dubath et al., 2005). \\
A complete INTEGRAL orbit lasts ${\sim}$3 days, but scientific data cannot 
be accumulated when the instrument is crossing the radiation belts, reducing 
the useful observing time ${\sim}$ 2.5 days. The instrument in-flight
performance is given in Roques et al. (2003).


\section{Observations}

Each INTEGRAL orbital revolution consists of several exposures (typically, 30-40 minutes 
pointings dithering around the target). Due to the small 
number of detectors, imaging which relies on these 
observations in dithering mode (Jensen et al., 2003), the pointing 
direction varying by steps of $2^\circ$ within a $5 \times 5$ square or a 7-point hexagonal pattern.

The data used to perform this analysis were recorded from 2003, March 3 to 2003, 
October 19, including in total 42 revolutions (table 1). They cover the region
 $-50^\circ$ $\leq$ l $\leq$ $50^\circ$, $-25^\circ$ $\leq$ b $\leq$ 
$25^\circ.$ Most of these data have been obtained through the GCDE
(Galactic Center Deep Exposure) part of the Integral core program (Winkler et al., 2003).

Data polluted by solar flares and radiation belt entries are excluded.
After image analysis and cleaning, there is  2552 pointings, which represent  
5.75$\times$10$^6$ seconds of effective observing time.


\section{Data analysis}


The data analysis is done in 2 steps. The first one concerns the determination of the 
positions of the emitting sources. We have used SPIROS, a software delivered in the 
INTEGRAL OSA package to produce images. In a second step, we have developed
a dedicated software to extract fluxes from the  positions obtained with SPIROS.


\subsection{Image generation}


SPIROS V6 (SPI Iterative Removal of Sources) algorithm  has been  used to derive source positions, 
with the MCM background option (mode 5). \footnote{Software and manual are available on the ISDC 
site (http://isdc.unige.ch/).}
SPIROS (Skinner \& Connell, 2003), a part of the Integral Science Data Centre (ISDC) package 
produces synthetic and simplified sky images : 
they consist of a limited number of  sky pixels, those which contain excesses above a given threshold.
An important limitation is that sources are considered as constant
during the image reconstruction process.
As a consequence, at low energy ( $<$ 50 keV), some exposures 
exhibit an unacceptable  $\chi^{2}$ fit of raw data to the reconstructed sky image 
convolved by the response matrix along with biased residuals 
distribution. We found this effect to be due to intensity
variations of the most intense sources, which are not taken into
account.
To suppress the effect of the variability of the strongest sources,
a special iterative scheme described in \S 4.2.2 has been used.

SPIROS works both with known and unknown  sources. If an input catalog is given, SPIROS will
first build a sky model with the proposed sources then looks for a number a new
ones required by the data.
One must take care of the influence of the source position errors on
the image generation. If the source position introduced in SPIROS is
different from the true position (even slightly), the iterative removal of sources algorithm
leads to subsequent artefacts in the image.
This problem may become important for strong sources or when a large
number of sources has inaccurate positions.
Thus, whenever possible, the exact source positions should be
introduced in SPIROS as  {\it a-priori} knowledge (input catalog) in order to avoid
 these cumulative errors.
But, {\it a contrario}, to use an input catalog with too many (non-emitting) sources also leads to unstable
solution. Our analysis philosophy (see \S 5) takes into account both limitations.


\subsection{Time dependent sources and background fluxes determination algorithm (time-model-fit)}


A complete model fitting procedure (called "time-model-fit") based on the likelihood 
statistics has been developed (See Annexe A-1).
The input sky model can include both point sources and diffuse components.
The main features of the algorithm are: \\
- A self determination of the background distribution on the detector plane.\\
- A time dependent background normalization determination \\
- A time dependent flux determination  for each point source with its own timescale (meaningless
 for diffuse components).


\subsubsection{Background determination}


The SPI imaging system, although using a combination of a coded mask with a 
position sensitive detector, needs, due to the small number of
detector pixels, a dithering scheme to increase the
number of sky pixels that can be reconstructed. 
This results in a time modulation of the sky signal.
In order not to confuse such modulation with background variations, the background
versus  time profile of the 19 detectors has to be evaluated.
The construction of a time-dependend background model constitutes a key point of 
SPI data analysis. 

Analyses of ``empty-field'' observations have shown that, for any energy band, the relative 
count rates (uniformity map) of the 19 Ge detectors are constant while the global 
amplitude (normalization factor) varies with time. We thus have to determined the relative
 count-rates of the 19 Ge detectors (background counts-ratio pattern  
depending on the energy band) leaving normalization as the only free parameter for the 
background intensity. This is included inside the model fitting procedure as described in annexe A-2
with the background amplitude able to vary on the pointing ($\sim$2500 s) timescale.


\subsubsection{Taking into account source variability}


The SPI image reconstruction relies on the 
dithering. As a result, variability of sources has to be explicitly included 
in the system of equations to be solved. This is included in our model fitting procedure 
 by means of additionnal equations (Annexe A-3). For each source, the allowed variation 
 time scale can be chosen. In practice, it has to be carefully used as the number of
  parameters (unknown fluxes) 
necessary to describe the sky will increase, correspondingly the significances decrease.
There is also a mathematical limitation: since there are 19 independent data per pointing, 
it  means that, whatever the number of pointings, there can not be more than 19 variable sources on
the time scale of a pointing.
 
However, most of the sources are weak enough to neglect 
the influence of their variability on the image reconstruction, and thus can
be considered as constant. Even flaring sources are too weak and/or shortliving 
(e.g. SGR 1806-20)  to have any effect on long period.\\
In this work, the "time-model-fit" tool has been used for a dedicated variability treatment 
applied only to the most intense sources, namely, 4U 1700-377, SCO X-1 and OAO 1657-4154, below 50 keV. 
We build for them the light curves in a one  pointing time scale. The corresponding contributions
in the count space are then substracted from the original data to derive a "corrected" data set 
to use as  SPIROS input. This cleaning is done from the original data set for each iteration, as 
the strong sources contributions can be affected by new sources introduction.\\
Applying this scheme to the
whole data set in the 25-50 keV energy band improves  the global reduced
$\chi^{2}$ from 3.24 (standard procedure) to 1.33 (Table 3).


\section{Catalogue generation}


Images are built using all our data in the following energy bands : 
 300 - 600 keV, 150-300 keV, 50-150 keV and 25-50 keV. 
In order to minimise bias (see \S 4.1), the first step of the analysis was performed without
any {\it a-priori }
 information about known sources. The {\it a priori}  knowledge is introduced 
progressively.
For each image, SPIROS is parametrized  in order to search for up to 30 new excesses above
2$\sigma$ in addition to the input catalog built from the previous iteration. This process
(i.e. to fix found positions) suppresses the instabilities of such an IROS algorithm. Moreover,
such a procedure leads to a minimal sky model able to represent the data.

The catalog generation process is the following:
\begin{itemize}
\item The first image  is obtained from SPIROS with an empty input catalog and gives  a list of source
candidates.
\item  A catalog is built with: (i) All identified sources above 5$\sigma$   with their 
celestial positions. (ii) The unidentified excesses above 5$\sigma$
 with the positions found.  The source identification process is described \S 5.1.
 This current catalog is completely regenerated  at each iteration as significances can evolve.
 
\item We run SPIROS again with this catalog as input, obtain a new list of sources/excesses, and
restart with step 2
\item This iterative analysis continues until no other potential 
excess $\geq$ 2$\sigma$ can be found.
\end{itemize}

A flow chart illustrates this process in Figure 1.\\

The final position catalog for a given energy band is built from the last iteration,
where sources and excesses with a significance threshold fixed to 4 $\sigma$ are accepted.
The use of a dedicated catalog for each energy band allows us to restrict
the number of free parameters to the minimum needed, thus decreasing
the flux detection limit at high energy.
Figure 2 and 3 present the images obtained by SPIROS in the 25-50
keV and 50-150 keV energy bands using our corresponding catalogs in input.


\subsection{Source identification}


To obtain source identification X-ray source catalogs are used. Primarly, the
ISGRI catalogs (Bird et al., 2004, Revnivtsev et al., 2004) are used and, if nothing is  found, 
the ISDC catalog is used (Ebisawa et al., 2003 \footnote{ available at http://isdc.unige.ch}).

Each excess is considered associated with the nearest known emitting X/$\gamma$ ray
sources whose distance is less than $1^\circ$.
This rather high value, 3 times the theoretical value for a $5\sigma$ detection 
(Dubath et al., 2005), is used because the source localisation
precision is degraded when more than one source is present,
especially for crowded regions. Note however that this value is much less than the 
geometrical resolving power (2.6$^\circ$), as the dithering pattern amounts to  improving
the imaging system.\\
So, two known sources closer than $1^\circ$ can be retrieved depending on the statistics (Dubath et al., 2005).
 The Galactic Center region is an exception as
too many sources are present in a small area. So we represent the $1^\circ$ region around
the Galactic Center  by only one source, namely 1E 1740.7-2942.
 
A more sophisticated scheme is used for the 25-50 keV band: the
dataset used is divided in two parts, one constructed with odd pointing
numbers the other with even pointing numbers; then two sub-images are
built.  A non-identified excess is considered
as  source candidate if it is detected in both images
(cross validation method).  This method drastically reduces
the number of false source detections introduced by systematics,  mainly due to
variable sources.\\
The number of excesses as function of the threshold level is shown
in table 2. Above 150 keV and below 5$\sigma$, the number of unidentified
excesses increase rapidely. From this,  empirical thresholds of 5$\sigma$
and 4 $\sigma$ respectively below and above 150 keV have been fixed.


\subsection{Final flux extraction}

Once the  position catalogs completed, we build a complete sky model including the detected
point sources plus diffuse emission morphologies (described is \S 6.1).\\
Fluxes and significances for each of this sky model components are thus computed in 
 a set of energy bands, using "time-fit -model" (\S 4.2).
The fluxes of point sources and diffuse emission are assumed constant (except for the 20-50 keV band, where 
4U 1700-377, SCO X-1 and OAO 1657-4154 vary on the time scale of one pointing).
The background normalization is adjusted on a time scale of one pointing.\\ 
Excesses above 4.5$\sigma$ are kept to form the final catalogs. Their fluxes and significances are 
computed again to be consistent; thus the final catalogs contain some sources below 
4.5$\sigma$ that were previously detected above this  threshold. This process is
described in Figure 4.

The final $\chi^{2}$ obtained with this method between the proposed sky model and the data are shown in table 3.


\subsection{The catalog}


The use of a dedicated input catalog for each energy band allows us to
optimize the signal-to-noise ratio per energy band.\\ 
These catalogs are far from exhaustive:
In the computed 1-year-averaged emission, weak or short-transient sources can be 
completely washed out. Moreover, as we looked for a minimal sky model, and given the modest 
SPI angular resolution, only one source is necessary to represent the complex
regions (1E 1740.7-2942 for instance). 

This work, based on the whole dataset, has been complemented by applying
the same procedure in 3 subsets, sorting data according to the average 
Galactic longitude of each observation (table 1):  positive (l $>5^\circ$), 
negative (l $< -5^\circ$) and central ($-5^\circ$ $\leq$ l $\leq$ $5^\circ$) longitudes, 
that have more or less equal exposures. These new datasets allow us to
add four sources in the 25-50 keV band catalog. These sources are
labelled with an asterisk.

As the final result, in the 25-50 keV domain the catalog contains 63 excesses 
above 4$\sigma$; 59 have been identified with known hard X-ray objects in the IBIS catalog, 4 are tentatively
associated with X-ray sources (labelled with **).  
Among the tentatively identified sources, the source at (l,b)=(1.94$^\circ$,-2.02$^\circ$) 
has a quite high significance of 11.2$\sigma$, while its flux is only 6.3 mCrab.
It could represent several weak (eventually extended) sources below ISGRI 
detection threshold that are integrated within the SPI angular resolution.
It could also results from variable sources not taken into account or from
errors accumulation due, for example, to a wrong 
center-of-gravity of the many sources that compose our so-called `1E 1740.7-2942' source     
actually fixed at the 1E 1740.7-2942 position. Intensive simulations are needed to understand the
behaviour of the instrument in such a crowded region, and is beyond the scope of this paper. 

Mean fluxes and significances are
given in 3 energy bands: 25-50 keV, 50-150 keV,150-300 keV (table 4).
In the 50-150 keV band, the catalog contains 20 excesses above 4$\sigma$ (table 5);
all but one correspond to confirmed sources, the last
identification is only tentative. Finally, 4 known objects emit
significantly above 150 keV (table 6).

\section {Spectral components of the Galactic ridge emission}

We will now concentrate on the
relative contribution of point sources to the Galatic ridge emission
as a function of the energy range.

In the soft $\gamma$ ray regime, multiple components contribute 
to the total emission. These include discrete sources, positron annihilation 
line and three-photon positronium annihilation continuum, and a diffuse continuum
resulting probably from cosmic ray interactions with the interstellar medium.

SPI is sensitive to both discrete sources and diffuse emission. As a consequence the discrete sources should always
be extracted simultaneously with diffuse emission. However it is                                             
difficult to separate/distinguish extended and discrete sources
emission if  too many free parameters are to be determined: the error
bars then become very large, and it may be impossible to derive meaningful 
information from the analysis.
Therefore  as much {\it a priori} information as possible is included in the analysis:\\
- precise external source locations is used in catalogs.\\
- spatial morphologies derived from previous works for each
diffuse component are used and fixed {\it a priori}.\\

\subsection{Sky model and components of soft $\gamma$-ray Galactic emission}

\subsubsection{Point sources model versus energy}

In order to estimate the total point source emission, it is obvious that all the discrete sources 
(also weak, transient and possibly undetected) should be included 
in the analysis. 
On the other hand, increasing the
number of sources decreases the significance of the measurement. Thus only the 
significant sources are introduced in the analysis,  their number depending on the 
 considered energy band. Three spectral regions are considered:\\
\begin{itemize}
  
\item In the 20-150 keV energy range, the 63 sources of the 25-50 keV
catalog have been introduced.
\item In the 150-300 keV energy range, the 20 sources of our 50-150
keV catalog is used.
\item Above 300 keV, the 4 sources detected above 150 keV have been considered.
\end{itemize}

\subsubsection{CR diffuse continuum}

The diffuse emission morphology  in the 50-400 keV 
estimated by measurements from OSSE/GRO 
(Purcell et al., 1996, Kinzer et al., 1999) 
and confirmed with subsequent simultaneous RXTE and OSSE observations 
(Valinia et al., 2000), is 
broadly distributed in longitude with a $5^\circ - 6^\circ$ FWHM in latitude 
and a 
$\sim \pm35^\circ$ extent in longitude. However since most of this emission at 
least in the 
low energy range is due to point sources (Lebrun et al., 2004), it cannot 
be used to 
represent the CR diffuse component. A better model may be the CO 
map (Dame et al., 2001) since it is a tracer of the interstellar matter.
 In the absence of a better model and for simplicity, the CR 
diffuse continuum spatial morphology is modelled by the CO map. 
Nevertheless, this point constitutes a weakness, as finally the morphology 
is poorly determined.

\subsubsection{Positron annihilation line and three-photon positronium continuum }

The annihilation line detected by SPI has been used to study the 511 keV 
spatial morphology (Kn\"odlseder et al., 2003). This component is equally
well described by models that represent the stellar bulge, by halo populations 
or by an azimuthally symmetric Gaussian 
with FWHM of $\sim$ $8^\circ$ (Kn\"odlseder et al., 2005). We thus modelled it  
with this last simpler hypothesis.\\
A recent study of the positronium annihilation emission with SPI shows that its
all-sky distribution is consistent with that of the 511 keV electron-positron annihilation
(Weidenspointmer et al., 2005).
In the absence of any other information, the three-photon 
positronium continuum  is assumed to have the same spatial distribution as the 
annihilation line. 

\subsection{Extraction of sky model component fluxes}
Component intensities of the sky model are extracted channel by channel with the time-model-fit 
algorithm described in \S 4.2. Then, count rates are converted into photon spectra for each component.
The total point source contribution to the Galactic emission is  obtained by summing all the point source 
spectra in the central radian, excluding those from Sco X-1 and Cen A which are located 
at high galactic latitude (b $\ge$ 20 $^\circ$).\\
 Figure 5 shows the composite spectrum of the three major components.

The summed point source emission (upper one) can be represented by a power law of index 
2.64$^{+0.12}_{-0.13}$ with an exponential cutoff of 504$^{+1100}_{-218}$ and a 100 keV
flux of 4.36$^{+0.96}_{-0.69}$ $\times$ 10$^{-4}$ photons~cm$^{-2}$~s$^{-1}$~keV$^{-1}$, for
a reduced $\chi^{2}$ of 1.74 for 13 degrees of freedom (dof).

The CR diffuse emission spectrum is more  complex. Below 50 keV it clearly exhibits a soft component.
In the 300-500 keV domain, there is clearly some cross-talk between
the positron annihilation  and the CR diffuse continuum components. This
can be easily understood as we try to fit simultaneously two sky model
distributions that have a common part in the inner Galaxy where the
exposure time is the highest. \\
The CR diffuse continuum fit above 50 keV, excluding 300-500 keV points (figure 5), 
gives a photon power law index of $\sim$ -1.8 $^{+0.5}_{-0.4}$  and a 100 keV flux 
of (6 $\pm$2) $\times$ 10$^{-5}$ photons~cm$^{-2}$~s$^{-1}$~keV$^{-1}$ for a reduced $\chi^{2}$ 
of 0.3(8 dof).\\
 In addition, a fit using only data whose pointing direction is outside the 
central radian, and supposed to contain insignificant electron-positron annihilation 
emission, gives a power index of 1.7$^{+0.5}_{-0.4}$ above 50 keV.\\
 These two indexes are
fully compatible with the
hard component expected from a `cosmic-ray interaction' model (Skibo et al., 1993), which can be approximated by
a power law with index of $\sim$ -1.65 over the 50 keV-10 MeV range. As we poorly constrain this parameter, we
 decided to fix the power law index to 1.65. \\
Finally, in the 20-1800 keV energy range, and excluding the 300-500 keV band,the CR diffuse 
continuum spectrum is  modelled by two components to give 
a reduced $\chi^{2}$ of 0.5 (8 dof): The first one is a  power law with a photon index 1.7 $^{+0.5}_{-0.4}$ 
and a cutoff energy of  20 keV, having a 50 keV flux of  
(3.1 $\pm$ 0.7) $\times$ 10$^{-4}$ photons~cm$^{-2}$~s$^{-1}$~keV$^{-1}$.
The second one is a power law with a photon index fixed to 1.65 and a 100 keV flux of
(4.8 $\pm$ 2.2) $\times$ 10$^{-5}$ photons~cm$^{-2}$~s$^{-1}$~keV$^{-1}$
that represents the CR diffuse above 50 keV up to several MeV (including OSSE and COMPTEL results).

From this study, the contribution of point sources to the total Galactic emission can  
be derived: 85\% in the 25-50 keV energy range, 90\% in the 50-110 keV
and 85\% in the 110-140 keV range. This fraction then
decreases as the positronium emission begins to be significant.\\
Second, the 511 keV flux extracted in a 10 keV width channel is
$\sim$ 0.9$\times$ 10$^{-3}$ photons~cm$^{-2}$~s$^{-1}$. This value is
compatible  with that obtained in
a completely different way by Kn\"odlseder et al. (2005).\\
 From these data, we also 
extract a positronium fraction (defined as in Brown \& Leventhal,
1987)  of $\sim$ 0.9 as a rough estimate. In fact, the CR diffuse component in the
300-500 keV energy range is clearly overestimated, leading to
an underestimation of the positronium flux.
A more complex and adequate method has thus been proposed for this spectral region (see next section).\\

\subsection{Complete photon model fitting (above 300 keV)}

The photon spectra extraction method applied above (\S 6.2) presents the advantages to be 
spectral model independent and relatively easy to implement. Its "simplicity" lies on approximations
well-suited only for energies below $\sim$ 300 keV. Indeed, 
 the SPI response is averaged over all the exposures for a given
component,  and the energy redistribution matrix is not properly taken
into account. The latter can lead to significant  errors particularly in the
case of photon spectra with positive slope, such as the positronium continuum (i.e. above 300 keV). 
 
For these reasons  a complete model fitting software has been developed,
in which the emission photon spectral models of each  sky components are
convolved with the full SPI response (Sturner et al., 2003) for each pointing. This
algorithm is detailed in Annex B.
In the other hand, this kind of algorithm allows  to fix the shape of the 
spectra as a function of the energy, via an analytique emission model. 
As a consequence, the problem of ``cross-talk'' previously encountered 
is drastically reduced.\\
Each photon spectral component follows  an appropriate emission model whose parameters 
can be adjusted simultaneously to the whole set of
data along with the background intensity. 
Given the size of the dataset such an algorithm can lead to
prohibitive computing time and memory space if the number of
free parameters is too high.
Fortunately, the benefit of this method is more significant at high energy, 
where the source number is low.

Moreover, we limit the number of free parameters by fixing some of
them. First, as the used energy band is limited to 300 keV- 1 MeV, 
the slope of the power law that modelled the CR diffuse  emission is fixed to 1.65 while the flux is to be fit. 
Second, the point source emission consists of 4 discrete sources (table 4) whose spectra are
represented a by power law with an exponential cut-off.  
Because the source statistics are low, the indices and normalizations of these
power laws have been previously determined below 150 keV  by the
first method (\S 6.2), while the cutoff energy is to be fitted.
Results may slightly depend  on the sources spectral shape assumed above
300 keV.\\ 
The annihilation spectrum (positronium continuum plus line) is then determined
by the complete model fitting.

A line flux of (0.93$\pm$0.15) 10$^{-3}$ photons~cm$^{-2}$~s$^{-1}$ and a
positronium fraction of 0.97$^{+0.09}_{-0.07}$ are found. The 100 keV flux of the 
diffuse emission is 5$\pm$3 $\times$ 10$^{-5}$ photons~cm$^{-2}$~s$^{-1}$~keV$^{-1}$,
in perfect agreement with  previous  
determinations. The reduced $\chi^{2}$ of the fit is 0.99 (630337 dof), the reduced $\chi^{2}$
for each energy band are distributed between 0.95 and 1.2.\\
The result is illustrated in figure 6.


\section{Discussion and conclusions}


Despite the SPI instrument's moderate angular resolution, and benefitting from the  
large number of pointings, a first catalog of hard
X-ray sources detected by SPI/INTEGRAL in the Galactic plane between
$-50^\circ$ $\leq$ l $\leq$ $50^\circ$, $-25^\circ$ $\leq$ b $\leq$ 
$25^\circ$ has been derived.\\
This catalog contains 63 sources and gives their fluxes up to 300 keV. It
has been built by means of model-fitting procedure simultaneously applied 
to all pointings of the whole dataset of 5.7$\times$10$^6$ seconds, and represents 
the minimal averaged sky model needed to decribe the data.
Thus transient sources could escape detection. Moreover, due to the modest SPI angular
resolution, our sky model in crowded regions is simplified, with one point
source representing the combined emission of all sources near 1E 1740.7-2942. 
On the other hand, this catalog is perfectly suited to estimate the point source
contribution to the Galactic ridge emission.\\
The diffuse emission components (continuum, 511 keV and positronium) have been
studied by fixing their spatial distributions, as a first order model, and taking into account
our catalog. \\
The main conclusions are :
\begin{itemize}
\item Point sources contribute at least  for 85-90\% of the total Galactic emission in the 
energy range 25 to 140 keV. This independantly confirms the ISGRI result that point 
sources largely dominate the total Galactic emission (Lebrun et al., 2004) at least up to 250 keV.
The interstellar emission is around 15 \% of the total emission up to 150 keV. 
The low energy part ($<$ 250 keV) of the total spectrum is source dominated 
while the high energy part ($>$ 250 keV) is diffuse dominated. 

\item The CR
 diffuse emission (50 - 250 keV) can be fitted  with a power law joining smoothly the high
energy continuum as measured by OSSE (Kinzer et al., 1999)  and COMPTEL (Strong et al., 1994) above 1 MeV.

\item No additional spectral component is needed above the  power law
spectrum to model the diffuse continuum above 50 keV.
The soft diffuse component reported by Kinzer et al. (1999) in this region is mostly due to
point sources not taken into account by this instrument.

\item Below 50 keV, a real diffuse soft component can still exist, but
      may also
be  related to very steep/weak X-ray sources not included in the
present analysis.

\item There is no evidence of 511 keV point source emission, in agreement
      with the conclusion drawn by Kn\"odlseder et al. (2005).

\item A positronium fraction 0.97$^{+0.09}_{-0.07}$  and a 511 keV flux of
      (0.93 $\pm$0.15) 10$^{-3}$ photons~cm$^{-2}$~s$^{-1}$. This positronium
      fraction is compatible with the OSSE determination: 0.93 $\pm$ 0.04
(Kinzer et al., 2001) and SPI/INTEGRAL determinations of 0.94 $\pm$ 0.06 
(Churazov et al., 2005).

\end{itemize}

These results constitute a first step as they are based
on SPI data taken during the first year of
operation and will be obviously refined in the future. 
An alternative approach is proposed by Strong et al. (2005).

The catalogs are available electronically as ApJ `supplementary electronic materials'.

\section*{Acknowledgments}

  The SPI/INTEGRAL project has been completed under the responsibility and leadership of CNES. We are grateful to ASI, CEA, CNES, DLR, ESA, INTA, NASA and OSTC for support.


\begin{deluxetable}{cccccc}
\tablewidth{0pt}
\tablecaption{
INTEGRAL/SPI observations used in the analysis. It covers GCDE 
cycle 1 and 2 and includes relevant public data. This
corresponds to a total of 2552 pointings for an effective time of 5754.2 
kiloseconds with a mean longitude l$_{mean}$=$0.6^\circ$ and latitude 
b$_{mean}$=$1.6^\circ$.}
\tabletypesize{\scriptsize}
\tablehead{
\colhead{Revolution} 
&\colhead{Start date}
&\colhead{End date} 
&\colhead{Useful duration }
&\colhead{l$_{mean}$ }
&\colhead{b$_{mean}$ } \\
\colhead{no} &\colhead{YY/MM/DD- hh:mn} &\colhead{YY/MM/DD- hh:mn} 
&\colhead{(kiloseconds)} &\colhead{(degrees)} &\colhead{(degrees) } }

\startdata
 47 & 2003/03/03 - 03:32 & 2003/03/05 - 18:02 &  136.0 &  -27.4 &    0.3 \\
 49 & 2003/03/09 - 03:00 & 2003/03/11 - 14:43 &   46.8 &  -20.7 &    1.1 \\
 50 & 2003/03/12 - 02:49 & 2003/03/14 - 16:44 &   92.1 &  -27.9 &    0.8 \\
 51 & 2003/03/15 - 02:34 & 2003/03/17 - 16:30 &   81.0 &  -12.1 &    0.8 \\
 52 & 2003/03/18 - 02:19 & 2003/03/20 - 16:36 &   91.6 &  -14.0 &   -0.1 \\
 53 & 2003/03/21 - 02:04 & 2003/03/23 - 15:43 &  143.2 &   -3.1 &    0.2 \\
 54 & 2003/03/24 - 01:50 & 2003/03/26 - 15:56 &  116.7 &  -15.8 &   -1.7 \\
 55 & 2003/03/27 - 01:35 & 2003/03/29 - 15:50 &   26.5 &  -10.3 &   10.7 \\
 56 & 2003/03/30 - 01:22 & 2003/04/01 - 15:27 &   49.6 &    4.1 &    4.0 \\
 57 & 2003/04/02 - 01:10 & 2003/04/04 - 15:10 &    0.7 &   -0.1 &   -1.1 \\
 58 & 2003/04/05 - 00:59 & 2003/04/07 - 14:20 &   18.1 &   -5.2 &   -2.9 \\
 59 & 2003/04/08 - 00:49 & 2003/04/10 - 14:44 &   48.7 &   12.7 &    1.1 \\
 60 & 2003/04/11 - 00:34 & 2003/04/13 - 14:59 &   43.6 &   -5.7 &   -0.4 \\
 61 & 2003/04/14 - 00:20 & 2003/04/16 - 14:17 &   51.2 &   -0.6 &    0.7 \\
 62 & 2003/04/17 - 00:06 & 2003/04/19 - 13:35 &   22.8 &    6.1 &    4.4 \\
 64 & 2003/04/22 - 23:37 & 2003/04/25 - 13:41 &  189.9 &   18.1 &   -0.1 \\
 65 & 2003/04/25 - 23:20 & 2003/04/28 - 13:11 &  201.7 &   24.2 &    0.1 \\
 66 & 2003/04/28 - 23:06 & 2003/05/01 - 13:25 &  145.0 &   26.2 &   -2.5 \\
 97 & 2003/07/30 - 16:59 & 2003/08/02 - 05:57 &  180.0 &   -1.4 &   23.7 \\
100 & 2003/08/08 - 15:18 & 2003/08/11 - 05:07 &  137.3 &  -18.7 &    0.6 \\
101 & 2003/08/11 - 15:57 & 2003/08/14 - 05:19 &  181.7 &   -2.1 &   23.5 \\
103 & 2003/08/17 - 14:39 & 2003/08/20 - 04:32 &  119.2 &   -7.8 &    0.6 \\
104 & 2003/08/20 - 14:25 & 2003/08/23 - 04:30 &  156.2 &  -42.1 &   31.1 \\
105 & 2003/08/23 - 15:06 & 2003/08/26 - 03:41 &  177.0 &   -0.1 &   -0.6 \\
106 & 2003/08/26 - 14:51 & 2003/08/29 - 03:38 &  166.7 &   -0.7 &    0.6 \\
107 & 2003/08/29 - 14:35 & 2003/09/01 - 03:17 &  197.6 &    0.8 &    0.8 \\
108 & 2003/09/01 - 14:22 & 2003/09/04 - 02:54 &  201.5 &    1.5 &   -0.0 \\
109 & 2003/09/04 - 14:07 & 2003/09/07 - 03:05 &  193.5 &   10.1 &   -0.8 \\
110 & 2003/09/07 - 13:52 & 2003/09/10 - 02:46 &  139.2 &    0.1 &   -0.6 \\
111 & 2003/09/10 - 13:39 & 2003/09/13 - 02:33 &  199.3 &    0.5 &   -0.3 \\
112 & 2003/09/13 - 13:26 & 2003/09/16 - 02:21 &  195.8 &    0.3 &   -0.3 \\
113 & 2003/09/16 - 13:14 & 2003/09/19 - 02:09 &  171.7 &   -0.1 &   -0.3 \\
114 & 2003/09/19 - 13:07 & 2003/09/22 - 01:58 &  177.3 &   -0.2 &   -0.3 \\
115 & 2003/09/22 - 12:55 & 2003/09/25 - 02:01 &  193.3 &    0.0 &   -0.2 \\
116 & 2003/09/25 - 11:47 & 2003/09/28 - 01:41 &  148.5 &  -14.8 &   -2.0 \\
117 & 2003/09/28 - 11:35 & 2003/10/01 - 01:50 &  188.5 &   22.0 &    0.9 \\
118 & 2003/10/01 - 11:21 & 2003/10/04 - 01:45 &  180.8 &  -19.2 &   -1.2 \\
119 & 2003/10/04 - 11:08 & 2003/10/07 - 01:12 &  178.9 &   -1.9 &   -1.3 \\
120 & 2003/10/07 - 10:55 & 2003/10/10 - 01:02 &  190.6 &    2.6 &    0.1 \\
121 & 2003/10/10 - 10:45 & 2003/10/13 - 00:46 &  198.9 &   14.3 &    0.8 \\
122 & 2003/10/13 - 10:35 & 2003/10/16 - 00:32 &  199.0 &   13.6 &    0.5 \\
123 & 2003/10/16 - 10:25 & 2003/10/19 - 00:22 &  176.3 &   26.9 &    0.3 \\
\enddata

\end{deluxetable}

\begin{deluxetable}{lcccc}
\tablewidth{0pt}
\tablecaption{Number of execesses detected in the 25-50 keV, 50-150 keV, 150-300 keV and
300-600 keV energy bands.} 

\tabletypesize{\scriptsize}
\tablehead{
\colhead{Significance}
&\colhead{25-50 keV} 
&\colhead{50-150 keV} 
&\colhead{150-300 keV} 
&\colhead{300-600 keV} \\
\colhead{$\sigma$} &\colhead{(All images)}  &  &\colhead{} }
\startdata

7$\sigma$    & 50(47)     & 16(15)    &     2(2)       &  1(1)       \\
6$\sigma$    & 56(53)     & 20(19)    &     4(4)       &  3(3)       \\
5$\sigma$    & 62(58)     & 24(19)    &     4(4)       &  4(3)       \\
4$\sigma$    & 69(64)     & 30(21)    &     7(4)       &  11(5)      \\

\enddata
\tablecomments{Before subtracting diffuse and annihilation radiation.
The number of excesses above 2 $\sigma$ are respectively 105, 42, 17 and 16 
for the 25-50 keV, 50-150 keV, 150-300 keV and 300-600 keV. The 25-50 keV result is the
combination of 4 images (l $>5^\circ$, l $< -5^\circ$, $-5^\circ$ $\leq$ l $\leq$ $5^\circ$ and the sum, see \S 5.3). 
In parenthesis is indicated the number of excesses associated with an
ISGRI detected sources (Revnivtsev et al., 2004, Bird et al., 2004).}
\end{deluxetable}

\begin{deluxetable}{lcc}
\tablewidth{0pt}
\tablecaption{Reduced $\chi^{2}$ between the sky image model and data.} 

\tabletypesize{\scriptsize}
\tablehead{
\colhead{Energy band}
&\colhead{Reduced} 
&\colhead{Degree of}  \\
\colhead{keV} &\colhead{$\chi^{2}$}   &\colhead{freedom} }
\startdata

  300 - ~600 &   1.027 & 45867  \\
  150 - ~300 &   1.013 & 45869  \\
  ~50 - ~150 &   1.125 & 45855  \\
 ~25 - ~50   &   3.242 & 45809 \\ 
 ~25 - ~50*  &   1.328 & 45806  \\

\enddata
\tablecomments{(*) ScoX-1, 4U 1700-377 and OAO1657-4154 contributions have been removed from the data.}

\end{deluxetable}

\begin{deluxetable}{rlccccc}
\tablewidth{0pt}
\tablecaption{Source catalog and fluxes in the 25-50 keV, 50-150 keV and 150-300 keV energy bands.} 

\tabletypesize{\scriptsize}
\tablehead{
\colhead{No} 
&\colhead{Name}
&\colhead{l} 
&\colhead{b} 
&\colhead{25-50 keV} 
&\colhead{50-150 keV} 
&\colhead{150-300 keV} \\
&&\colhead{deg} &\colhead{deg} &\colhead{mCrab} &\colhead{mCrab} &\colhead{mCrab} }

\startdata

~1 &                 Cen A &   -50.48 &    19.38 &   54.7 $\pm$    2.9 (  18.9) &   72.7 $\pm$    6.3 (  11.5) &   85.4 $\pm$   21.8 (   3.9)\\
~2 &           4U 1516-569 &   -37.88 &     0.04 &   18.0 $\pm$    3.0 (   6.0) &   15.0 $\pm$    6.6 (   2.3) &  $<$  46.1                  \\
~3 &           4U 1538-522 &   -32.58 &     2.18 &    9.6 $\pm$    2.0 (   4.9) &  $<$   9.2                   &  $<$  33.5                  \\
~4 &           4U 1608-522 &   -29.06 &    -0.86 &   17.8 $\pm$    1.6 (  11.0) &   17.8 $\pm$    3.9 (   4.6) &  $<$  28.5                  \\
~5 &           4U 1636-536 &   -27.08 &    -4.81 &   19.2 $\pm$    1.5 (  12.8) &   12.5 $\pm$    3.6 (   3.5) &  $<$  26.1                  \\ 
~6 &       IGRJ16167-4957* &   -26.92 &     0.51 &  5.8   $\pm$    1.4 (   4.3) &  $<$   7.2                   &  $<$  26.7                  \\
~7 &       IGR J16318-4848 &   -24.37 &    -0.44 &   20.9 $\pm$    1.5 (  13.7) &  $<$   7.4                   &  $<$  27.5                  \\
~8 &            4U 1630-47 &   -23.08 &     0.24 &   76.1 $\pm$    1.5 (  50.1) &   49.8 $\pm$    3.7 (  13.6) &  $<$  26.9                  \\
~9 &              GX 339-4 &   -21.06 &    -4.32 &    8.6 $\pm$    1.2 (   7.0) &   11.8 $\pm$    2.9 (   4.1) &  $<$  20.5                  \\
10 &       IGR J16418-4532 &   -20.81 &     0.52 &   14.1 $\pm$    1.2 (  11.4) &    9.6 $\pm$    2.9 (   3.3) &  $<$  21.2                  \\
11 &           4U 1705-440 &   -16.67 &    -2.33 &   28.3 $\pm$    1.2 (  22.9) &   18.1 $\pm$    2.9 (   6.3) &  $<$  21.1                  \\
12 &           4U 1702-429 &   -16.11 &    -1.31 &    8.4 $\pm$    1.3 (   6.5) &    6.9 $\pm$    3.1 (   2.2) &  $<$  22.5                  \\
13 &         OAO 1657-4154 &   -15.63 &     0.32 &   83.0 $\pm$   1.7    (50.3)        &   34.1 $\pm$    2.3 (  14.7) &  $<$  17.1          \\
14 &           4U 1735-444 &   -13.94 &    -6.99 &    6.8 $\pm$    0.9 (   7.3) &  $<$   4.2                   &  $<$  15.1                  \\
15 &       IGR J17195-4100 &   -13.03 &    -2.12 &   11.1 $\pm$    0.8 (  13.9) &    7.1 $\pm$    1.9 (   3.7) &  $<$  14.1                  \\
16 &           4U 1700-377 &   -12.25 &     2.18 &  164.1 $\pm$   1.4   (119.7)        &   75.8 $\pm$    2.0 (  38.3) &   16.6 $\pm$    7.4 (   2.2)\\
17 &              GX 349+2 &   -10.88 &     2.76 &   14.2 $\pm$    1.2 (  11.4) &  $<$   6.1                   &  $<$  22.9                  \\
18 &      IGR J17252-3616* &    -8.49 &    -0.36 &   11.8 $\pm$    1.4 (   8.5) &  $<$   3.2                   &  $<$  11.8                  \\
19 &            4U 1746-37 &    -6.48 &    -4.97 &    7.7 $\pm$    0.6 (  13.0) &    6.7 $\pm$    1.4 (   4.7) &  $<$  10.6                  \\
20 &              GX 354-0 &    -5.69 &    -0.15 &   23.0 $\pm$    0.6 (  37.2) &    7.5 $\pm$    1.5 (   4.9) &  $<$  11.3                  \\
21 &          GRS 1724-308 &    -3.67 &     2.31 &   20.3 $\pm$    0.5 (  38.7) &   12.0 $\pm$    1.4 (   8.9) &  $<$  10.1                  \\
22 &           4U 1822-371 &    -3.14 &   -11.29 &   29.6 $\pm$    0.8 (  38.1) &    9.6 $\pm$    1.8 (   5.2) &  $<$  13.4                  \\
23 &       IGR J17488-3253 &    -3.03 &    -2.64 &   15.4 $\pm$    0.9 (  16.9) &    9.1 $\pm$    2.3 (   4.0) &  $<$  16.9                  \\
24 &       IGR J17464-3213 &    -2.74 &    -1.83 &    6.2 $\pm$    1.1 (   5.7) &    9.3 $\pm$    2.7 (   3.5) &  $<$  19.6                  \\
25 &          GRS 1734-292 &    -1.11 &     1.41 &    7.8 $\pm$    0.6 (  12.3) &  $<$   3.4                   &   15.6 $\pm$    6.2 (   2.5)\\
26 &               Sco X-1 &    -0.90 &    23.77 &  216.8 $\pm$   1.6   (136.6)        &   20.9 $\pm$    2.3 (   9.2) &   18.9 $\pm$    8.5 (   2.2)\\
27 &        1E 1740.7-2942 &    -0.87 &    -0.10 &   80.4 $\pm$    1.1 (  71.8) &   78.9 $\pm$    2.7 (  28.8) &   45.6 $\pm$   10.0 (   4.6)\\
28 &          SLX 1744-299 &    -0.72 &    -0.89 &   25.6 $\pm$    1.2 (  20.6) &  $<$   6.3                   &   24.7 $\pm$   11.5 (   2.1)\\
29 &              V2400Oph &    -0.15 &     8.73 &   14.2 $\pm$    0.9 (  16.4) &    4.8 $\pm$    2.1 (   2.3) &  $<$  15.2                  \\
30 &       EXMSB 1709-232* &     0.59 &     9.27 &   16.8 $\pm$    3.4 (   4.9) &  $<$   4.2                   &  $<$  15.2                  \\
31 &          SLX 1735-269 &     0.81 &     2.41 &   10.1 $\pm$    0.5 (  18.7) &    3.8 $\pm$    1.4 (   2.8) &  $<$  10.0                  \\
32 &           RX J1832-33 &     1.54 &   -11.36 &    4.1 $\pm$    0.7 (   5.9) &    3.5 $\pm$    1.7 (   2.1) &  $<$  12.1                  \\
33 &         XTE J1807-294 &     1.93 &    -4.27 &    5.0 $\pm$    0.6 (   9.1) &  $<$   2.8                   &  $<$  10.4                  \\
34 &     AX J1758.0-2818** &     1.94 &    -2.02 &    6.3 $\pm$    0.6 (  11.2) &  $<$   3.0                   &   14.9 $\pm$    5.7 (   2.6)\\
35 &                GX 1+4 &     1.95 &     4.80 &   12.0 $\pm$    0.5 (  22.2) &    7.3 $\pm$    1.3 (   5.5) &  $<$  10.0                  \\
36 &                GX 3+1 &     2.30 &     0.80 &   10.5 $\pm$    0.6 (  17.6) &  $<$   3.1                   &  $<$  11.5                  \\
37 &           4U 1820-303 &     2.79 &    -7.92 &   14.8 $\pm$    0.6 (  25.1) &  $<$   2.8                   &   11.8 $\pm$    5.2 (   2.3)\\
38 &          GRS 1758-258 &     4.51 &    -1.36 &   61.2 $\pm$    1.0 (  59.9) &   76.5 $\pm$    2.5 (  30.7) &   59.6 $\pm$    9.2 (   6.5)\\
39 &            4U 1849-31 &     4.97 &   -14.35 &    4.3 $\pm$    1.0 (   4.5) &  $<$   4.3                   &  $<$  14.9                  \\
40 &                GX 5-1 &     5.09 &    -1.01 &   20.6 $\pm$    1.1 (  19.4) &    6.7 $\pm$    2.6 (   2.6) &   19.2 $\pm$    9.5 (   2.0)\\
41 &    SAX J1810.8-2609** &     5.20 &    -3.43 &    2.5 $\pm$    0.5 (   4.8) &  $<$   2.6                   &  $<$   9.7                  \\
42 &       IGR J17597-2201 &     7.58 &     0.77 &    9.5 $\pm$    0.6 (  15.4) &  $<$   3.0                   &  $<$  11.2                  \\
43 &           4U 1745-203 &     7.72 &     3.81 &    4.3 $\pm$    0.6 (   7.7) &  $<$   2.7                   &  $<$  10.1                  \\
44 &                GX 9+9 &     8.53 &     9.04 &    4.5 $\pm$    0.7 (   6.8) &    3.3 $\pm$    1.6 (   2.1) &  $<$  11.3                  \\
45 &            GS 1826-24 &     9.28 &    -6.08 &   80.5 $\pm$    0.6 ( 125.9) &   54.0 $\pm$    1.5 (  35.8) &   30.7 $\pm$    5.6 (   5.5)\\
46 &           SGR 1806-20 &    10.01 &    -0.21 &    5.8 $\pm$    0.7 (   8.8) &  $<$   3.2                   &  $<$  11.9                  \\
47 &    AX J1812.2-1842/SNR 011.2-00.3** &    11.90 &    -0.16 &  3.7 $\pm$   0.7  (5.3)                 &    6.4 $\pm$    1.7 (   3.8) &  $<$  12.7                  \\
48 &          PKS 1830-211 &    12.15 &    -5.72 &    4.8 $\pm$    0.7 (   7.0) &    7.6 $\pm$    1.6 (   4.6) &  $<$  12.3                  \\
49 &               GX 13+1 &    13.52 &     0.12 &    7.9 $\pm$    0.8 (  10.2) &    4.7 $\pm$    1.9 (   2.5) &  $<$  14.0                  \\
50 &               GX 17+2 &    16.44 &     1.28 &   17.1 $\pm$    0.8 (  21.1) &  $<$   3.9                   &  $<$  14.2                  \\
51 &             M 1812-12 &    18.03 &     2.40 &   20.0 $\pm$    0.9 (  22.7) &   16.1 $\pm$    2.1 (   7.7) &  $<$  15.3                  \\
52 &     AX J1825.1-1253** &    18.49 &    -0.14 &    3.5 $\pm$    0.9 (   4.0) &  $<$   4.3                   &  $<$  15.9                  \\
53 &       SNR 021.5-00.9* &    21.51 &    -0.88 &    5.4 $\pm$    0.9 (   6.0) &  $<$   4.8                   &  $<$  17.5                  \\
54 &      IGR J18325-0756* &    23.71 &     0.57 &    6.2 $\pm$    0.9 (   6.8) &    6.3 $\pm$    2.5 (   2.5) &  $<$  18.9                  \\
55 &           4U 1850-087 &    25.36 &    -4.32 &    5.7 $\pm$    1.1 (   5.2) &  $<$   5.2                   &  $<$  19.0                  \\
56 &                Kes 73 &    27.39 &     0.01 &    5.1 $\pm$    1.1 (   4.5) &    8.1 $\pm$    2.8 (   2.9) &  $<$  20.5                  \\
57 &         XTE J1855-026 &    31.08 &    -2.09 &   17.9 $\pm$    1.4 (  13.0) &   10.6 $\pm$    3.3 (   3.3) &  $<$  23.8                  \\
58 &              1916-053 &    31.35 &    -8.49 &    6.4 $\pm$    1.6 (   4.0) &  $<$   7.3                   &   31.9 $\pm$   13.3 (   2.4)\\
59 &       AX J1852.6+0038 &    33.65 &     0.02 &    6.2 $\pm$    1.6 (   3.9) &  $<$   7.6                   &  $<$  28.0                  \\
60 &            4U 1901+03 &    37.19 &    -1.25 &   22.9 $\pm$    2.2 (  10.6) &  $<$  10.4                   &  $<$  37.9                  \\
61 &                SS 433 &    39.70 &    -2.24 &   13.6 $\pm$    2.8 (   4.8) &  $<$  13.2                   &  $<$  47.6                  \\
62 &           4U 1909+07* &    41.90 &    -0.81 &   13.5 $\pm$    3.3 (   4.2) &  $<$  17.7                   &  $<$  59.4                  \\
63 &          GRS 1915+105 &    45.37 &    -0.21 &   91.6 $\pm$   15.4 (   6.0) &  $<$  36.9                   &  $<$  97.4                  \\

\enddata

\tablecomments{ {\bf 1 Crab = 0.1433, 0.0889 and 0.0178 ph/s respectively in the
25-50, 50-150 and 150-300 keV energy bands} \\
(*) :Source detected in the negative or
  positive longitude image only.
The values per energy band are the flux and error bar with significance in parenthesis or
2$\sigma$ upper limit when the source is not seen above 2$\sigma$. 
The values reported there take into account diffuse emission and annihilation radiation continuum.\\
(**) : Sources tentively identified.}
\end{deluxetable}

\begin{deluxetable}{lccc}
\tablewidth{0pt}
\tablecaption{50-150 keV source catalog.} 

\tabletypesize{\scriptsize}
\tablehead{
\colhead{l}
&\colhead{b} 
&\colhead{Name} 
&\colhead{50-150 keV} \\
\colhead{deg} &\colhead{deg}  &&\colhead{mCrab} }
\startdata

                 Cen A &   -50.48 &    19.38 &   76.8 $\pm$    6.7 (  11.5)\\
           4U 1608-522 &   -29.06 &    -0.86 &   19.5 $\pm$    3.9 (   5.0)\\
            4U 1630-47 &   -23.08 &     0.24 &   55.5 $\pm$    3.2 (  17.6)\\
         OAO 1657-4154 &   -15.63 &     0.32 &   37.0 $\pm$    2.0 (  18.2)\\
           4U 1705-440 &   -16.67 &    -2.33 &   22.5 $\pm$    2.2 (  10.2)\\
           4U 1700-377 &   -12.25 &     2.18 &   80.2 $\pm$    1.6 (  48.8)\\
            4U 1746-37 &    -6.48 &    -4.97 &    9.4 $\pm$    1.3 (   7.2)\\
              GX 354-0 &    -5.69 &    -0.15 &    5.3 $\pm$    1.3 (   4.2)\\
          GRS 1724-308 &    -3.67 &     2.31 &   11.1 $\pm$    1.2 (   9.0)\\
           4U 1822-371 &    -3.14 &   -11.29 &    9.3 $\pm$    1.7 (   5.4)\\
       IGR J17488-3253 &    -3.03 &    -2.64 &   10.2 $\pm$    1.9 (   5.3)\\
       IGR J17464-3213 &    -2.74 &    -1.83 &    8.2 $\pm$    1.9 (   4.4)\\
               Sco X-1 &    -0.90 &    23.77 &   22.4 $\pm$    2.4 (   9.3)\\
        1E 1740.7-2942 &    -0.87 &    -0.10 &   87.3 $\pm$    1.3 (  67.1)\\
                GX 1+4 &     1.95 &     4.80 &    6.1 $\pm$    1.2 (   5.0)\\
          GRS 1758-258 &     4.51 &    -1.36 &   82.9 $\pm$    1.2 (  67.8)\\
            GS 1826-24 &     9.28 &    -6.08 &   57.0 $\pm$    1.5 (  38.4)\\
AX J1812.2-1842/SNR 011.2-00.3** &    11.90 &    -0.16 &    5.7 $\pm$    1.5 (   3.9)\\
          PKS 1830-211 &    12.15 &    -5.72 &    8.0 $\pm$    1.5 (   5.2)\\
             M 1812-12 &    18.03 &     2.40 &   16.9 $\pm$    2.0 (   8.6)\\

\enddata

\tablecomments{The catalog contains only the sources detected
in the 50-150 keV band. The counts for diffuse continuum and sources 
were extracted simultaneously to obtain these data.(**) Source tentatively identified.} 
\end{deluxetable}

\begin{deluxetable}{lccc}
\tablewidth{0pt}
\tablecaption{150-300 keV source catalog.} 

\tabletypesize{\scriptsize}
\tablehead{
\colhead{l}
&\colhead{b} 
&\colhead{Name} 
&\colhead{150-300 keV} \\
\colhead{deg} &\colhead{deg}  &&\colhead{mCrab} }
\startdata

  -2.74 &    -1.83 &    IGR J17464-3213 &   21.2 $\pm$    4.0 (5.3) \\
  -0.87 &    -0.10 &     1E 1740.7-2942 &   61.4 $\pm$    4.1 (14.8) \\
   4.51 &    -1.36 &       GRS 1758-258 &   75.8 $\pm$    4.0 (18.9) \\ 
   9.28 &    -6.08 &         GS 1826-24 &   28.6 $\pm$    4.3 (6.6) \\


\enddata

\tablecomments{The catalog contains only the sources detected above
5 $\sigma$ in the 150-300 keV. The counts for diffuse continuum, 
three-photon positronium continuum, 511 keV line and sources 
were extracted simultaneously to obtain these data.} 
\end{deluxetable}






\clearpage
\begin{figure}

\plotone{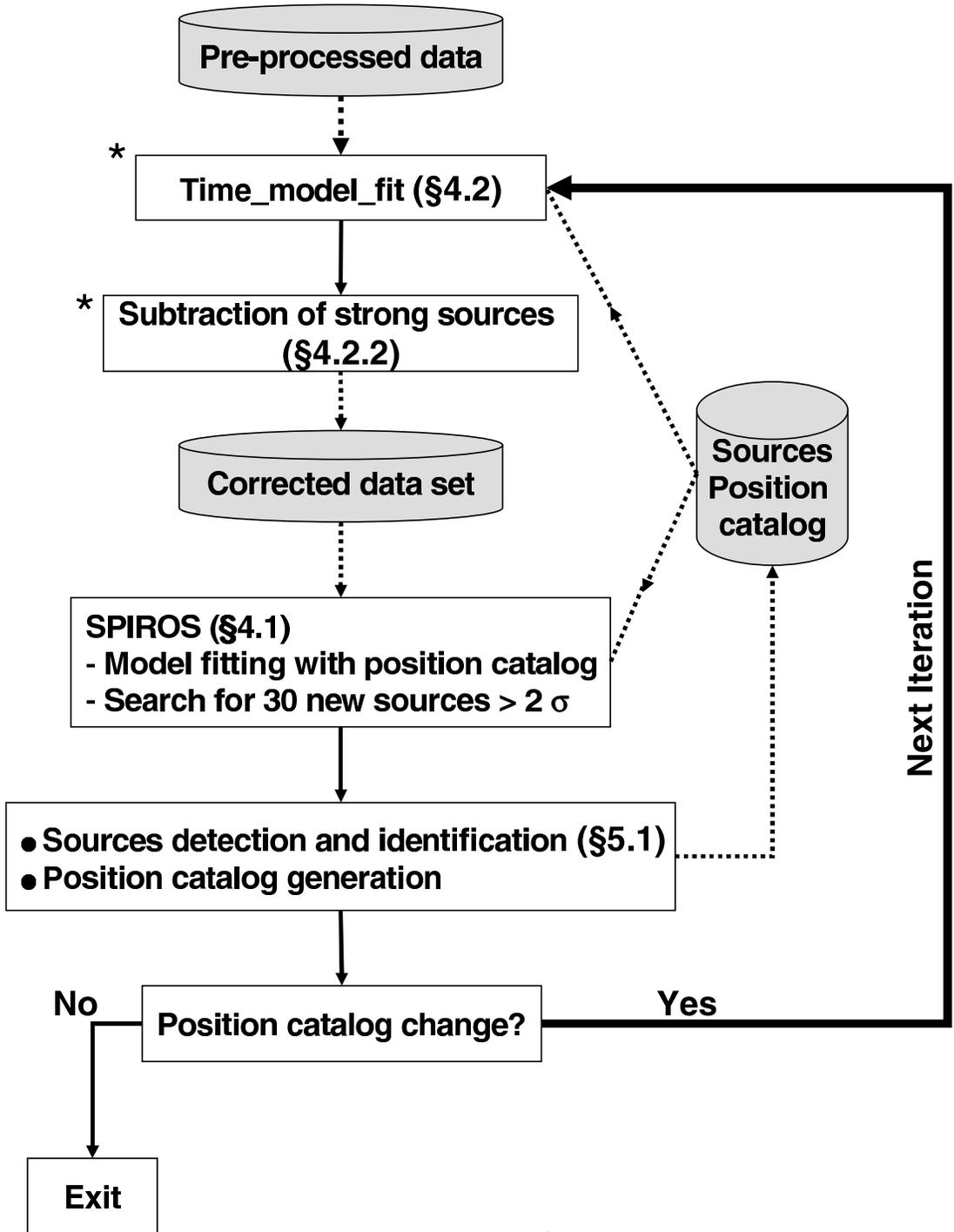}
\caption{Source positions catalog generation flowchart.}
\end{figure}

\clearpage
\begin{figure}
\vspace{-8cm}
\plotone{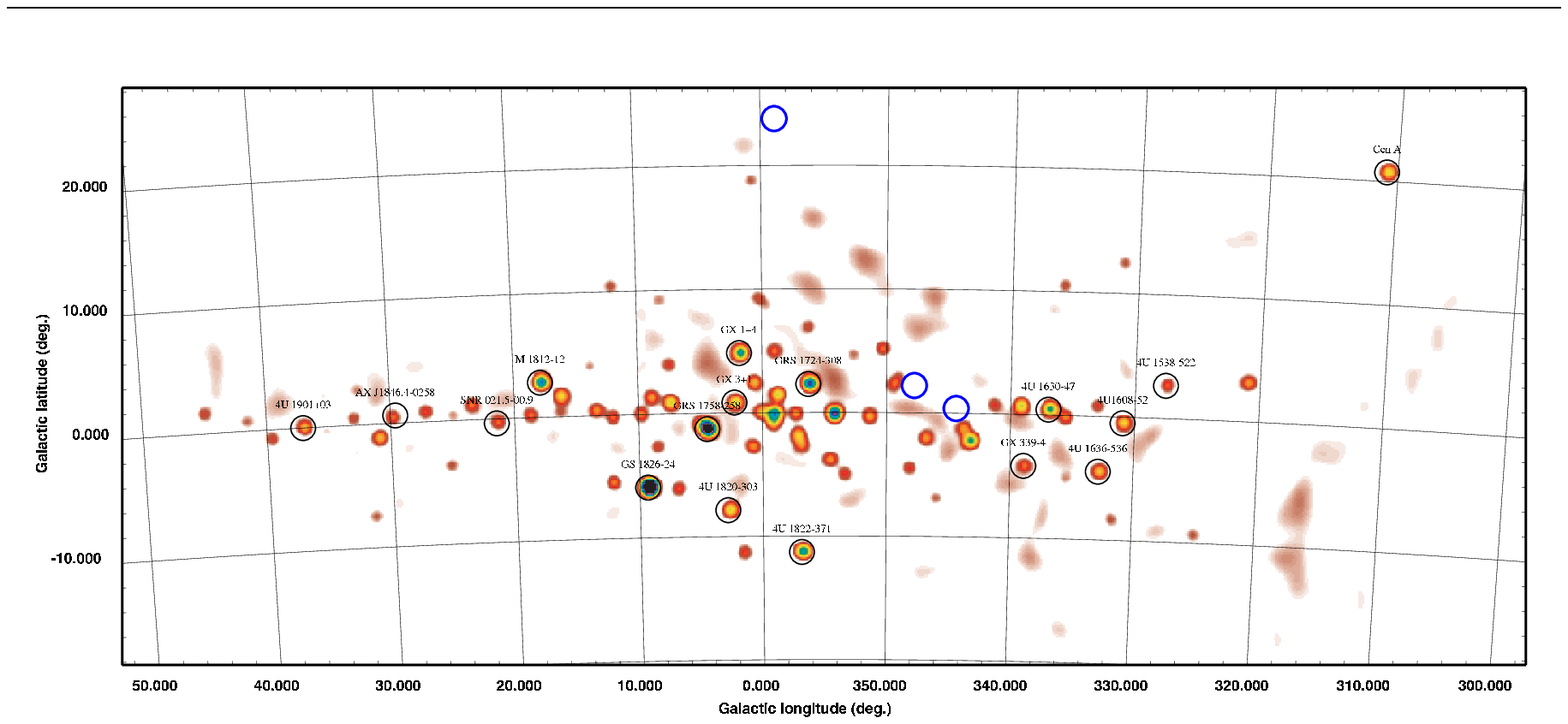}
\vspace{-8cm}
\caption{25-50 keV significance image of the Galactic Center region. 
Lower Threshold is 1$\sigma$ while upper one is fixed to 50$\sigma$.  Some sources position are marked with a circle of radius $1^\circ$.
Contributions of 4U 1700, Sco X-1 and OAO 1657-4154 have
been removed from the data, the blue circles indicate their positions.}
\end{figure}

\clearpage
\begin{figure}
\vspace{-8cm}
\plotone{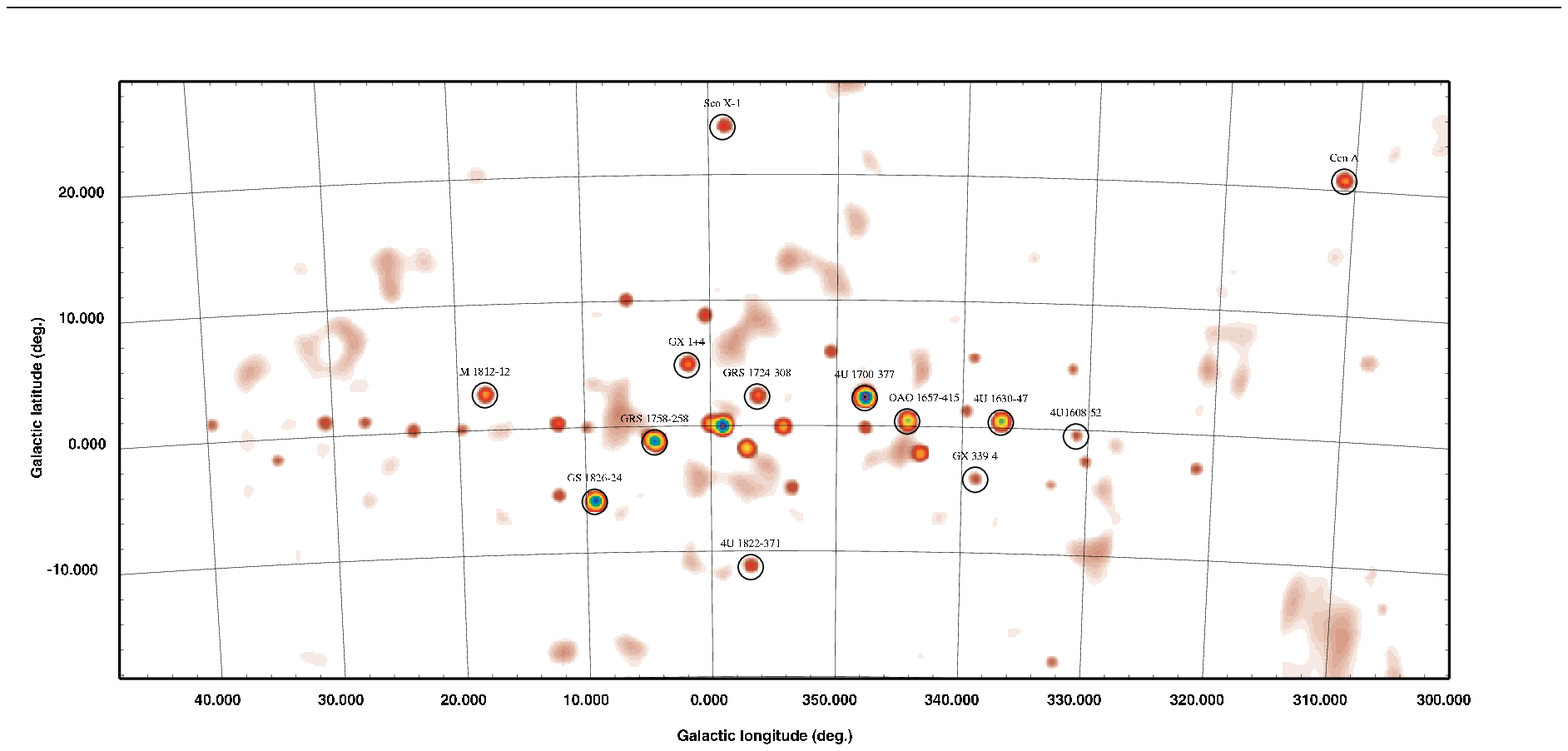}
\vspace{-8cm}
\caption{50-150 keV significance image of the Galactic Center region. 
Lower Threshold is 0$\sigma$. Some sources position are marked with a circle of radius $1^\circ$.}
\end{figure}

\clearpage
\begin{figure}

\plotone{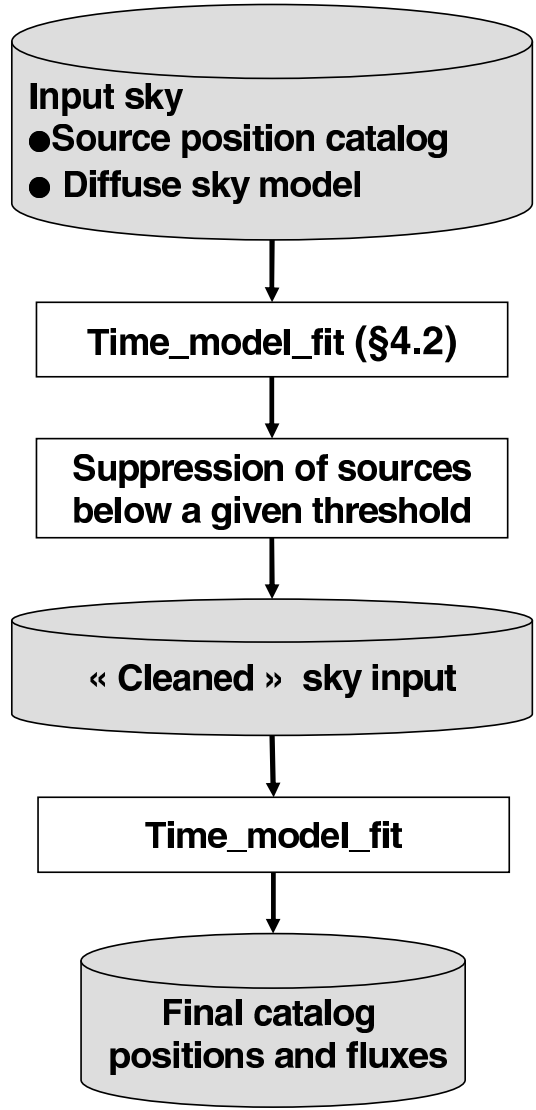}
\caption{Flux extraction algorithm flowchart.}
\end{figure}

\clearpage
\begin{figure}
\plotone{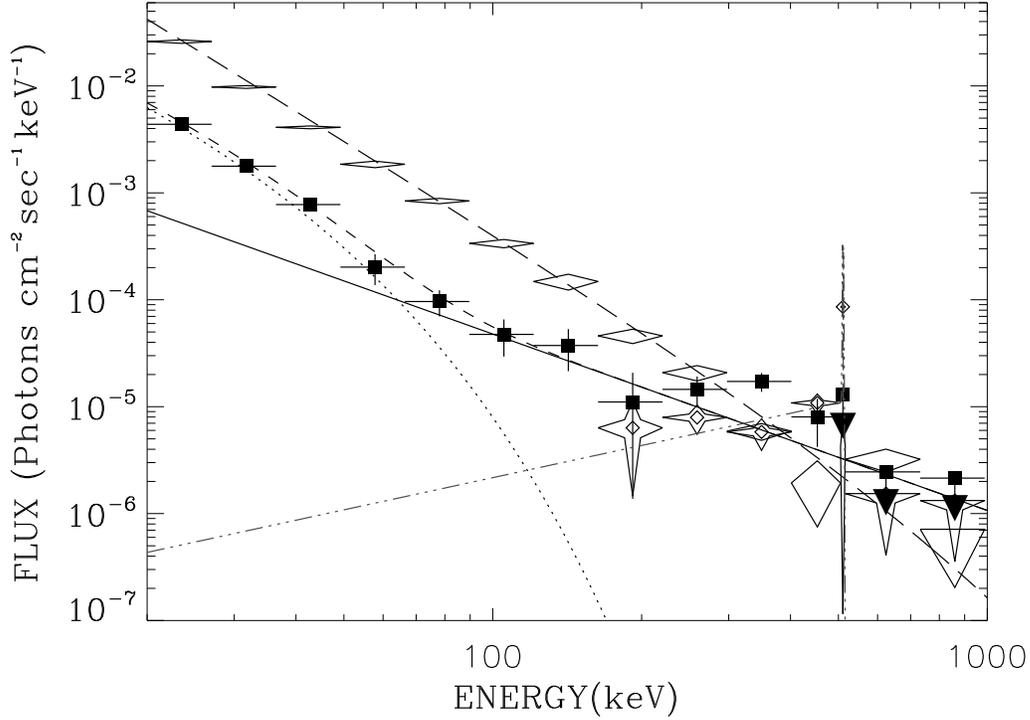}
\vspace {-10cm}
\caption{Combined spectra obtained by the channel by channel method 
described in \S 6.2. There are 15 energy bins logarithmically spaced, 
and  a bin of 10 keV centered at 511 keV.
Spectra below 150 keV are obtained using the 63 sources catalog (table 4), 
150-300 keV spectra are obtained using the 20 sources catalog (table 5) 
and above 300 keV using the 4 sources catalog (table 6). CR diffuse 
continuum  spectrum (fill squares) and its modelisation (dashed line) that 
consits of two components : a power law of index 1.65 (solid line), and the power law plus 
exponential cutoff (dot line) are shown.  
Summed point sources emission (diamonds) and its fits (long-dash line),
and the approximation of annihilation 
radiation spectrum (star) and  its fit (dash-dot-dot line) are also
represented.}
         
\end{figure}

\clearpage
\begin{figure}
\plotone{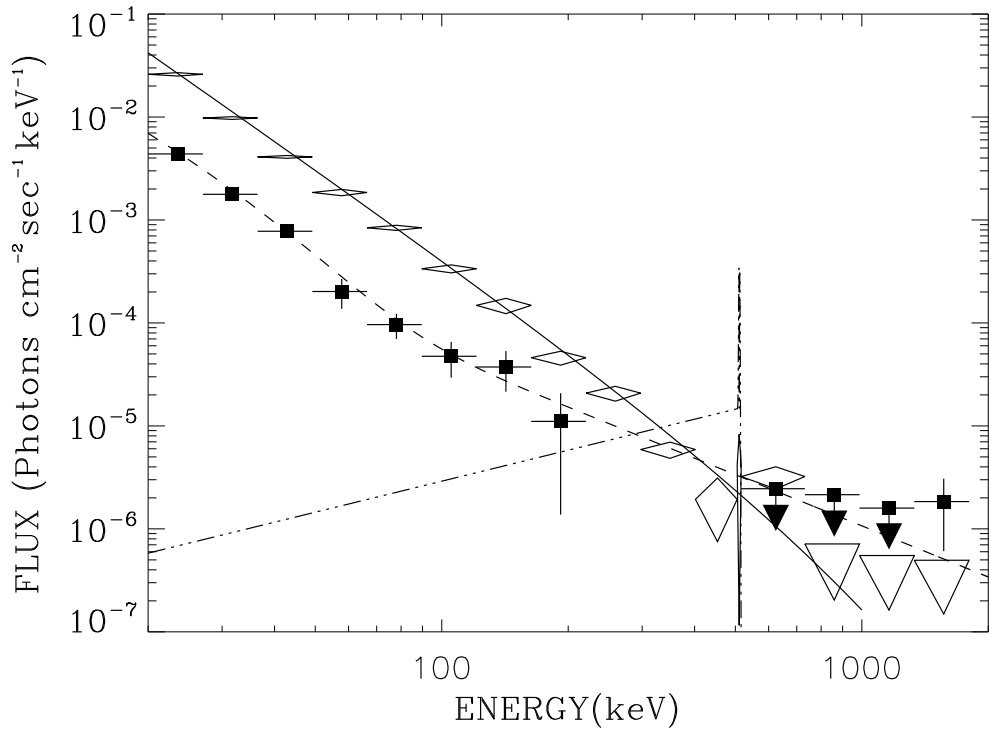}
\vspace {-10cm}
\caption{Combined spectra  obtained by the model fitting method
  described in \S 6.3 above 300 keV. Spectra below 300 keV are
  obtained as in fig 3. Summed point sources emission (diamonds) and its fit (upper solid line), 
  CR diffuse continuum (fill squares) and annihilation radiation spectrum fit (Dash dot dot line)
  are shown.  }
\end{figure}

\clearpage

\appendix
\section{Algorithms}

\subsection{Maximum likelihood algorithm}

The relation between the expected data, instrument aperture response, background and the $N_s$ sources which light the detectors is

\begin{displaymath}
E(d,p,e)=\sum_{j=1}^{N_s} M(d,p,e,\theta_j,\phi_j) t(d,p) I(\theta_j,\phi_j,e)+B(d,p,e)
\end{displaymath}
  
where E(d,p,e) and B(d,p,e) are respectively the expected data and background
in the energy bin e, pointing p and detector d. $I(\theta_j,\phi_j,e)$ is the
intensity of the source number j, or sky image  pixel j, in the direction 
$(\theta_j,\phi_j)$ in the energy bin e. $M(d,p,e,\theta_j,\phi_j)$ the instrument response of the data element (d,p,e) to the sky image pixel j in the direction 
$(\theta_j,\phi_j)$. t(d,p) is the duration exposure for the detector d and 
pointing p.

The background term is rewritten as 

\begin{displaymath}
B(d,p,e)=t(d,p)U(d,e)
\end{displaymath} 

U(d,e) is the uniformity map that fixed the relative counts ratio between
the 19 Ge detectors (\S4.1). If the background amplitude varies on the pointing 
timescale, then an amplitude coefficient A(p) is introduced
to describe the variable background $B_v$, 

\begin{displaymath}
B_v(d,p,e)=t(d,p)U(d,e)A(p)
\end{displaymath} 

For a given energy bin e and to simplify the equation representation,
the sky image pixel direction  $(\theta_j,\phi_j)$ is represent by the sky 
pixel number j. The equation reduces to

\begin{displaymath}
E(d,p)=\sum_{j=1}^{N_s} M(d,p,j) t(d,p) I(j)+ U(d) t(d,p) A(p)
\end{displaymath}

For the maximum likelihood algorithm, the cash statistics (Cash, 1979)
is used and the following equation is to be maximized with respect to the parameters I and A (which are constrained to be positive):

\begin{displaymath}
L(I_1,...,I_N;A_1,..A_{N_p})=E(d,p)-N(d,p)log(E(d,p))
\end{displaymath}

where N(d,p) are the measured data (for energy bin e)
and $N_P$ the number of pointing.

\subsection{Background uniformity map estimation}

The background uniformity estimation is obtained in two steps.
First an uniformity map $\widetilde{U}(d)$ is assumed and the likelihood is optimized  with respect to I and A to give the best estimation 
$\widetilde{I}$ and $\widetilde{A}$. Second, knowing the current best estimate of the sources intensity and current background. The background map is obtained through the minimization for each detector of  
 
\begin{displaymath}
 \sum_{p=1}^{N_P}  \frac{\big( N(d,p)-S(d,p)-\beta (d) B(d,p) \big) ^2}{\sigma^2(d,p)}
\end{displaymath}

where S(d,p) is the total source counts contribution, B the current background and $\beta(d)$ a multiplicative factor of order 1 of the current background map for detector d.

\begin{displaymath}
\begin{array}{l}
S(d,p)=\sum_{j=1}^{N_s} M(d,p,j) t(d,p) \widetilde{I}(j) \\
B(d,p)=U(d) t(d,p) \widetilde{A}(p)   \\
\sigma(d,p)^2=E(d,p)  \\
\end{array}
\end{displaymath}

The minimization with respect to $\beta(d)$ gives

\begin{displaymath}
\beta(d)= \frac {\sum_{p=1}^{N_P} \big(N(d,p)-S(d,p) \big) B(d,p) / \sigma^2(d,p)}
{\sum_{p=1}^{N_P} B^2(d,p) / \sigma^2(d,p)}
\end{displaymath}

The best uniformity map is then
\begin{displaymath}
U(d)=\widetilde{U}(d) \beta(d)
\end{displaymath}

This process is repeated several times until the equivalent 
$\chi^{2}$ stops to decrease , the equivalent $\chi^{2}$ being defined as

\begin{displaymath}
\chi_{eqv.}^2=\sum_{d,p}  \frac{\Big( E(d,p)-D(d,p)) \Big) ^2}{E(d,p)}
\end{displaymath}

\subsection{Variable source}

The expected counts E from a given source k counts is

\begin{displaymath}
E_{k}(d,p)=M(d,p,k) t(d,p) I(k)
\end{displaymath}

For a variable source this equation is to be expanded in several equations, corresponding
to intervals whic the source is considered as constant

\begin{displaymath}
E_{k}(d,p) = \left\{ \begin{array}{ll}
M(d,p,k) t(d,p) I_1(k) & \textrm{if~}  p=1,\cdots,p_1 \\
M(d,p,k) t(d,p) I_2(k) & \textrm{if~}  p=p_1+1,\cdots, p_2\\
\cdots & \cdots \\
M(d,p,k) t(d,p) I_L(k) & \textrm{if~} p=p_{L-1}+1,\cdots,P \\
\end{array} \right.
\end{displaymath}

Finally, instead of a single intensity for source k, 
there is L intensities $I_1(k),\cdots,I_L(k)$ to be determined.

\section{Photon model-fitting}

The function to be minimized is 
\begin{displaymath}
\chi^2 =  \sum_{e=1}^{N_E} \sum_{p=1}^{N_P} \sum_{d=1}^{N_D} \frac{\big( N(d,p,e)-E(d,p,e) \big) ^2}{\sigma^2(d,p,e)}
\end{displaymath}

where $N_E$ is the number of detector energy bin (counts space), $N_P$ the number of pointing and $N_D$ the number of detector. N(d,p,e) and E(d,p,e) are respectively measured and expected counts for detector d, pointing p and detector energy bin e.
The following equation relies the response matrix, sources photon spectra, background and expected counts

\begin{displaymath}
\sum_{j=1}^{N_s} \sum_{e_{ph}=1}^{N_{E_{ph}}} R(d,p,j,E,e_{ph})t (d,p) F(j,e_{ph}) + B(d,p,e) = N(d,p,e)
\end{displaymath}

where $R(d,p,j,E,E_{ph})$ is the  response (aperture response and detector
redistribution matrix) of the data element (d,p,e) to 1 incident photon
in the sky direction  $(\theta_j,\phi_j)$ for the incident in the energy
bin energy $e_{ph}$.$N_s$ and  $N_{E_{ph}}$ are respectively the number of sources and
energy bin in the photon space. $F(j,e_{ph})$ the incident photon spectrum in the energy bin $E_{ph}$ (photon space) for 
the source j.

For each one of the $N_s$ sources, the incident photon spectrum is described by 
few parameters described by the vector $\vec \theta_j$

\begin{displaymath}
F_{j,e_{ph}}(\vec \theta_j) \equiv F(j,e_{ph}) \qquad j=1,\cdots,N_{s}; e_{ph}=1, \cdots, N_{E_{ph}}
\end{displaymath}

The background will depends on the assumed photon spectra (\S B.1) and thus on the parameters
set of parameters $\theta $ 

\begin{displaymath}
B_{d,p,e}( \theta ) \equiv B(d,p,e)  \qquad \textrm{with} \qquad
\theta  \equiv ( \vec {\theta_1};\vec {\theta_2}, \cdots, \vec {\theta_{N_s}} ) 
\end{displaymath}

The set of parameters $\theta $ are  adjusted to minimized the function $f(\theta) $ 
( $ \equiv \chi^2 $ ) through a non-linear minimization method. 

\subsection{The background}

Given the parameters which describes the photon spectra and thus the
photon spectrum of each sources  $\widetilde{F}(j,e_{ph})$, the background is
determined in each energy bin e. Given the best current background pattern
$\widetilde{U}(d)$, the background is modeled as 

\begin{displaymath}
B(d,p)=t(d,p) \widetilde{U}(d) A(p)
\end{displaymath}

The current best background amplitude $\widetilde{A}(p)$ is determined exposure 
by exposure through the minimization with respect to A(p) of

\begin{displaymath}
\sum_{d=1}^{N_D}  \frac{\big( N(d,p)-S(d,p)-A(p) t(d,p) \widetilde{U}(d) A(p) \big) ^2}{\sigma^2(d,p)}
\end{displaymath}

where $N_D$ is the number of detector and
\begin{displaymath}
\begin{array}{l}
S(d,p)=\sum_{j=1}^{N_s} \sum_{e_{ph}=1}^{N_{E_{ph}}} R(d,p,j,E,e_{ph}) (d,p) \widetilde{F}(j,e_{ph}) \\ 
\sigma^2(d,p) = N(d,p,e)
\end{array}
\end{displaymath}

\subsection{background uniformity pattern}

Having the sources parameters best estimate, the background uniformity pattern 
U(d) can be optimized as in \S{A.2}

\end{document}